\documentclass[5p,twocolumn,number,sort&compress]{elsarticle}            

\makeatletter
\def\ps@pprintTitle{%
  \let\@oddhead\@empty
  \let\@evenhead\@empty
  \def\@oddfoot{\reset@font\hfil\thepage\hfil}
  \let\@evenfoot\@oddfoot
}
\makeatother

\usepackage{amsmath}
\usepackage{amsfonts}
\usepackage{amssymb}

\newcommand{\eq}[1]{\begin{equation}
                     \begin{split} #1 \end{split}
                     \end{equation}}

\begin{document}


 
\title{Towards Universal Axion Inflation and Reheating in String Theory}

\author[rb]{Ralph Blumenhagen} 
\author[ep1,ep2]{Erik Plauschinn}

\address[rb]{Max-Planck-Institut f\"ur Physik, F\"ohringer Ring 6, 80805 M\"unchen, Germany \\[3pt]}
\address[ep1]{Dipartimento di Fisica e Astronomia ``Galileo Galilei'', Universit\`a  di Padova, Via Marzolo 8, 35131 Padova, Italy \\[3pt]}
\address[ep2]{
INFN, Sezione di Padova, Via Marzolo 8, 35131 Padova, Italy
}

 
\begin{abstract}
\noindent
The recent BICEP2 measurements of B-modes indicate a large
tensor-to-scalar ratio in inflationary cosmology, which 
points towards  trans-Planckian evolution of the inflaton.
We propose possible string-theory realizations thereof.  
Schemes for natural
and axion monodromy inflation are presented 
in the framework of the type IIB large volume scenario. 
The inflaton in both cases is given by the universal axion and its
potential is generated by F-terms.
Our models are shown to feature a  natural  mechanism for inflaton 
decay into  predominantly  Standard Model particles.
\end{abstract}


\maketitle


\section{Introduction}
\label{sec:intro}

The recent reports of the BICEP2 collaboration \cite{Ade:2014xna} indicate that 
for the first time there have been direct measurements of B-modes,
which are CMB imprints of primordial gravity waves.
Indeed, BICEP2 found the tensor-to-scalar ratio to be rather 
large, $r=0.2$. Realizing this value in slow-roll inflationary cosmology
requires a motion of the inflaton over trans-Planckian distances
in field space. This is difficult to achieve in a UV complete
theory of quantum gravity, such as string theory, as one has 
to have control over many possible higher-order operators, which
can spoil the slow-roll property. This is known as the $\eta$-problem, which
is a challenge in general and becomes even stronger for trans-Planckian
evolutions.

As a matter of fact, most string or string-inspired models of inflation
give much lower values of $r$ and would, if BICEP2 is confirmed,
be ruled out. 
In order to achieve the required control over
higher-order corrections, one can take advantage of the perturbative 
shift symmetry of axions, which is only broken non-perturbatively, by fluxes,
or  by the presence of branes.
Interestingly, axions are ubiquitously present in string
compactifications. 
In the prototype model, the inflationary potential takes
the form $V(\theta)=\Lambda^4(1-\cos(\theta/f))$, where $f$ denotes the
instanton decay constant of the axion $\theta$. In fact, this simple 
model is known to lead to a large  tensor-to-scalar ratio, consistent
with all other cosmological parameters, if $f>M_{\rm pl}$.
It is a priori unclear whether in this  regime the  effective field-theory 
description is still trustable, however, 
this model of natural inflation \cite{Freese:1990rb,Adams:1992bn} 
fits the data amazingly  well.
Expanding the potential to quadratic order, it essentially reproduces
Linde's model of chaotic inflation \cite{Linde:1983gd},
which in fact has been shown to be compatible with large field variations in
\cite{Kaloper:2008fb,Kaloper:2011jz,Kaloper:2014zba}.

In order to avoid the regime $f>M_{\rm pl}$, extensions of this simple
axion model have been proposed. One 
is N-flation \cite{Kim:2004rp,Dimopoulos:2005ac}, where the collective
evolution of $N$ axions in a sort of radial direction 
leads to the same predictions, but where each individual axion 
only travels over a sub-Planckian distance and has a decay constant 
$\sqrt{N}f_i>M_{\rm pl}$.  
A second variation is so-called
 axion monodromy inflation \cite{Silverstein:2008sg,McAllister:2008hb,Palti:2014kza}, 
 where the shift symmetry is broken by the presence
of a brane,  inducing an approximately linear (or quadratic)
potential for the  axion. Thus, the former periodic axion is
``unwrapped'' and can now move over trans-Planckian distances, increasing
the energy by a certain amount each time it goes around the
period.

It turns out that  realizing
N-flation in a concrete string-theory model  is not an easy task
\cite{Grimm:2007hs,Cicoli:2014sva}. 
One generic problem with N-flation is that for $N\gg 1$ a
substantial renormalization of the Planck mass occurs.
Another  potential problem of
many  models is that at the end of the inflationary epoch, the
inflaton is not guaranteed to predominantly decay into
the visible sector, but the decay rates into a visible and a hidden sector
degree of freedom 
tend to be of the same order (see e.g. \cite{Cicoli:2010ha, Cicoli:2010yj}).

Clearly, in view of the BICEP2 results and its fairly constraining 
consequences,  it is important to study
what possibilities string theory can offer to realize such axion-inflation models. In this letter, we 
ignore the difficulties  the regime $f>M_{\rm pl}$ causes
and construct two models:  a string-theory model of natural inflation, 
and a new type of monodromy inflation, where
the shift symmetry of the axion is broken by fluxes instead 
of by the presence of branes.  
For concreteness, we focus on the LVS framework \cite{Balasubramanian:2005zx}
and consider as the inflaton (a linear combination involving)
the axionic component of the complex axio-dilaton  field
$S=C_0+i\exp(-\phi)$. Note that in most approaches before, this universal
axion was considered to be fixed at high mass scale by background
three-form fluxes.

We argue that type IIB string theory 
has the necessary ingredients to construct successful models
of inflation. In particular,
\begin{itemize}

\item{We assume  that the (flux) landscape admits points where 
the masses of the saxions (including the dilaton) are  hierarchically different
from the mass of $C_0$.
In particular, apart from the nearly massless axion of the big four-cycle in a LVS,
$C_0$ can be the lightest closed-string modulus, making it a good
candidate for the inflaton.}

\item{For natural inflation, the potential of the axion is generated
by non-perturbative effects from fluxed $E3$-instantons, whereas
for axion monodromy inflation the axion $C_0$
can appear   quadratically in the flux induced scalar potential.}

\item{There exists a mechanism guaranteeing that inflaton decay at the end 
of inflation predominantly goes into standard model (SM) degrees 
of freedom.}
\end{itemize}
This last point is one of  the very interesting aspects of the models
considered in this letter. Note furthermore that the relevant axion potentials
are  F-terms in an effective  spontaneously-broken supergravity theory,
which is in the same spirit as \cite{Marchesano:2014mla}.

Finally, note that an axion decay constant  $f>M_{\rm pl}$ corresponds to the non-perturbative (F-theory)
regime  $g_s>1$ of the type IIB superstring. We collect some indications
that the LVS scenario might be trustable even for string coupling
constants slightly larger than one, but of course conclusive
evidence requires the parametric control over infinitely many
perturbative corrections to the K\"ahler potential.


\section{Natural inflation}

Before we present our string-theory realization, let us recall
some results for the cosmological parameters derived from the simple
natural-inflation Lagrangian
\eq{
   {\cal L}= {1\over 2} \partial_\mu \theta\,  \partial^\mu \theta+
\Lambda^4\left(1-\cos\left({\textstyle {\theta\over f}}\right)\right) ,
}
where after canonical normalization the axion $\theta$ has a  period $2\pi f$. 
The slow-roll parameters 
\eq{
   \epsilon={M^2_{\rm pl}\over 2} \left({V'\over V}\right)^2\,, \hspace{40pt}
  \eta={M^2_{\rm pl}} \left({V''\over V}\right),
}
can be expanded in $\theta/f\ll 1$ as
\eq{
    \epsilon=\eta\sim {2M_{\rm pl}^2\over f^2} \left({f\over \theta}\right)^{2} 
\, .}
For this to be small one needs $M_{\rm pl}< \theta < f$ during inflation,
and inflation ends at $\theta_{\rm end}=\sqrt{2} M_{\rm pl}$.
The number of e-foldings are expressed as follows
\eq{
   N_e&={1\over M_{\rm pl}^2} \int^\theta_{\theta_{\rm end}} {V\over V'}\,
   d\theta\\
      &\sim{1\over 2 M_{\rm pl}^2} \int^\theta_{\theta_{\rm end}} \theta\, d\theta=
        {\theta^2\over 4 M_{\rm pl}^2} -{1\over 2}\, .
}
Thus, we write $N_e\sim {1\over 2\epsilon}$.
Therefore, for the spectral indices and the tensor-to-scalar ratio
one obtains the prediction
\eq{
     n_s&=1+2\eta-6\epsilon\sim 1-4\epsilon\sim 1-{2\over N_e}\,,\\
    n_t&=-2\epsilon\sim -{1\over N_e}\,,\\
    r&=16\epsilon\sim {8\over N_e}\, .
}
For 60 e-foldings, this rough  estimate  gives
\eq{
     n_s\sim 0.967\, ,\qquad  n_t\sim -0.017\, ,\qquad r=0.133\,,
}   
which is in good agreement with the recent measurements from Planck
for $n_s=0.9624\pm 0.0075$ and from BICEP2 for $r=0.20^{+0.07}_{-0.05}$.
The amplitude of the scalar power spectrum ${\cal P}=2.2\cdot 10^{-9}$ can be
written as follows
\eq{
   {\cal P}\sim {H^2_{\rm inf}\over 8\pi^2\epsilon M^2_{\rm pl}} \,,
}
leading to a Hubble constant during inflation of $H_{\inf}\sim 1.1\cdot
10^{14}\,$GeV. Via $V_{\rm inf}=3M_{\rm pl}^2\, H^2_{\inf}$ we can now
extract the mass scale of inflation as 
\raisebox{0pt}[0pt]{$M_{\rm inf}=V_{\rm inf}^{1/4}\sim2.1\cdot 10^{16}\,$}GeV, 
which is of the order of the GUT scale.
Finally, the mass of the axion 
$m^2_{\theta}=3\eta H^2_{\rm inf}$ comes out as $m_{\theta}\sim 1.7\cdot
10^{13}\,$GeV.

Note that since only the quadratic approximation of the cosine function
was used, these predictions are the same as for a quadratic potential,
which is nothing else than Linde's model of chaotic inflation.
The main problem to realize such a model in string theory
is the constraint $f>M_{\rm pl}$, which for all known cases
means that one is outside of the regime of validity of the
effective field theory approach. To overcome this restriction,
it was proposed to consider axions and break their shift symmetry by branes or fluxes,
generating a non-oscillatory potential for the axion and thus allowing it
to roll over trans-Planckian distances in field space, without
needing $\theta/f<1$. Hence, the periodic axion unfolds to allow
for non-trivial monodromies.


\section{Universal natural inflation in String theory}

Let us present a possible string-theory realization of the above scenario,
based on the well-established framework of moduli stabilization
in type IIB orientifolds on Calabi-Yau three-folds with $O7$- and $O3$-planes.
Recall that in the standard LVS one starts with a swiss-cheese 
type Calabi-Yau three-fold with K\"ahler potential
\eq{
\label{kaehlerpot}
  &{\mathcal K} = 
  \mathcal K_{\rm cs} (U) - \log(-i(S-\overline S))  \\
&\hspace{15pt}-2\log\left[ (T_b+\overline T_b)^{3\over2} -(T_s+\overline T_s)^{3\over 2}
    +\xi \left({\textstyle{S+\overline S\over 2}}\right)^{3\over 2}\,\right] ,
}
and fixes the axio-dilaton and complex structure moduli
via a tree-level three-form flux-induced superpotential 
$W=\int \Omega\wedge (F-i\hspace{1pt}S H)$.
For $\xi=0$, the no-scale structure leaves the K\"ahler moduli 
$T_b=\tau_b+i a_b$,  $T_s=\tau_s+i a_s$
massless at this level. Note that the masses of the complex structure moduli
and the axio-dilaton
are generically of the order $m_{\rm cs}\simeq 1/\mathcal V^2$, which is 
heavier than the K\"ahler moduli. Therefore, for constructing
a model with the universal axion as the inflaton, we need
to keep the axion essentially massless at the level of the flux-induced
potential.

Now, due to the landscape  of flux vacua, there could exist special but
still numerous non-supersymmetric vacua where the 
universal axion remains unfixed with its shift symmetry  still
intact. Therefore, the superpotential 
at the minimum can be independent of $C_0$.
For this letter, we leave aside the technical issue
of stabilizing e.g. the dilaton in such a way that its mass 
is hierarchical bigger than the mass of the axion.
Thus, 
we assume that the complex structure moduli and the dilaton 
are fixed either directly by fluxes, or by additional no-scale violating
terms, such that the above assumption is satisfied.
In this case the value of the superpotential $W_0$ at the minimum is a 
constant. Let us emphasize that a concrete realization of this
moduli stabilization scheme might be technically challenging and
 deserves a closer investigation \cite{Blumenhagen:future}.

In the LVS one  breaks the no-scale structure by a combination
of $\alpha'$-corrections leading to $\xi\ne 0$ and an  $E3$-instanton induced superpotential
\eq{
\label{superpot}
  W = W_{0} + A_s\,  e^{-\alpha_s T_s} \,,
}
where  $A_s=O(1)$ denotes the one-loop Pfaffian after integrating out the
complex structure moduli and $\alpha_s$ is a numerical factor.
Note that $A_s$ is independent of the axio-dilaton $S$.
For such an   instanton
to directly contribute to $W$ it has to have the right number
of fermionic zero modes. First, it has to be invariant of type $O(1)$ under
the orientifold projection. Second, sufficient conditions are that
it carries no deformation or Wilson line modulini and does 
not intersect any of the potentially present space-time
filling D7-branes. Del-Pezzo surfaces are good examples.

The main observation of \cite{Balasubramanian:2005zx} 
is that the scalar potential for the K\"ahler-moduli sector 
\eq{
          V=e^{\cal K}\, \left(\sum_a {\cal K}^{ab} D_a W D_b \overline W -3|W|^2 \right)
}
can be expanded in large volumes $\mathcal V$ of the compactification space.
With the superpotential \eqref{superpot}, one then finds
\eq{
\label{scalpota}
  V\simeq{\sqrt{\tau_s} \hspace{0.5pt} \alpha^2_s \hspace{0.5pt}|A_s|^2 e^{-2\alpha_s \tau_s}\over {\cal V}}
   -{{\tau_s} \alpha_s \bigl| A_s W_0 \bigr|  e^{-\alpha_s \tau_s} \over {\cal V}^2}
+\xi{ |W_0|^2 \over g_s^{3/2} {\cal V}^3} ,
}
admitting  a non-supersymmetric AdS minimum in which the moduli are stabilized
hierarchically as $\tau_b\gg \tau_s$. Concretely one finds
\eq{
      \tau_s\simeq {O(1)\over g_s}\, ,\hspace{25pt}
      a_s \simeq O(1)\,,\hspace{25pt}
     {\cal V}\sim {O(1)\over g_s^{1/ 2}}\, e^{\alpha_s \tau_s} \,,
}
and the masses of the moduli around this minimum scale with the
overall volume as
\eq{
\label{modulimass}
             m_{\tau_b}&\sim {M_{\rm pl}\over {\cal V}^{3\over 2}}\, ,\hspace{40pt}
              m_{a_b}\sim 0 \,, \\
              m_{\tau_s}&\sim {M_{\rm pl}\over {\cal V}}\, ,\hspace{40pt}
              m_{a_s}\sim {M_{\rm pl}\over {\cal V}} \, .
}
From the second line it follows that the small-cycle axion $a_s$ is not a good candidate
for axion inflation, as its mass is of the same
order as $m_{\tau_s}$ and actually heavier than $m_{\tau_b}$.
The big cycle axion $a_b$, after assuming  it can get a mass of order
$m_{a_b}\sim \exp(-{\cal V}^{2\over 3})$ via a non-perturbative 
effect, has been considered for axion inflation in \cite{Cicoli:2014sva}. 
To avoid the problem with $f>M_{\rm pl}$, more such big cycle
axions were introduced and a model of N-flation was successfully built.
However, let us mention that from the string-instanton zero-mode perspective 
it is questionable whether
 such a non-perturbative effect exist, as the big cycles tend
to come with many deformation and Wilson line modulini.

From now on, we  investigate  the axion $\theta:=C_0$ a bit closer.
First, computing the axion decay constant from the K\"ahler potential
we find from the $\log(S+\overline S)$ term in \eqref{kaehlerpot} that
\eq{
     f={M_{\rm pl}\over \sqrt 2}\, g_s\, ,
}
where the term in the second line of \eqref{kaehlerpot} gives
only a subleading correction.
Therefore, for  $f> M_{\rm pl}$ one needs $g_s>1$, 
which actually is outside the 
perturbative regime of string theory. 
We will come back to this delicate issue later.

At this stage the universal axion $\theta$  is still massless as it does not appear in the 
superpotential. However, in addition to standard $E3$-brane instantons also 
magnetized $E3$-brane instantons generically make contributions to $W$
(see e.g. \cite{Grimm:2011dj}).
Say as in \eqref{superpot}, we have an unfluxed $E3$-brane instanton wrapping
the small del-Pezzo divisordivsor $D_s$. Now, the  del-Pezzo surface $dP_r$
has $h_{11}=r+1$ holomorphic two-cycles.  Under the  orientifold
involutions these split into $h_{11}^+$ even and $h_{11}^-$ odd ones.
Turning on a background gauge flux $f$ along the odd cycles still preserves
the $O(1)$ property, and does not introduce modulini.
Therefore, such a configuration also makes a contribution to the
superpotential, so that one gets
\eq{
\label{superpotb}
  W = W_{0} + A_s\,  e^{-\alpha_s T_s} + B_s\,  e^{-\alpha_s (T_s+ S h(f))} +\ldots \,,
}
where the dots indicate that there can be many such contributions,
and where $h(f)={1\over 4\pi^2}\int_{D_s} f\wedge f\in \mathbb Z$.
If $h(f)$ is big enough, this second term does not affect the
moduli stabilization of e.g. $T_s$,  as it
is by a factor 
\eq{
\lambda_f=\exp\left(- \alpha_s {h(f)\over 2 g_s}\right)
}
smaller than the second term in \eqref{superpotb}.

Next, we want to compute the masses and the mixing of 
the two axions $\phi^{(1)}=a_s$ and $\phi^{(2)}=\theta$.
For this purpose \cite{Conlon:2007gk} we have to diagonalize the action 
\eq{
   {\cal L}_{\rm ax}= {\cal K}_{ab} \partial \phi^a \partial \phi^b -
                    V_{ab}  \phi^a \phi^b 
}
for the fluctuations
of these axions around their values at the LVS minimum of the
scalar potential.
Focusing on the order ${\cal V}$ and $\lambda_f$ dependence, we find
\eq{
         V_{ab}=\left(\begin{matrix} {O(1)\over {\cal V}^3} &
      {O(1)\, \lambda_f \over {\cal V}^3} \\[0.1cm]
     {O(1)\, \lambda_f \over {\cal V}^3} &
      {O(1)\, \lambda_f \over {\cal V}^3}
 \end{matrix}\right)
}
and
\eq{
         {\cal K}_{ab}=\left(\begin{matrix} {O(1)\over {\cal V}} &
      {O(1)\,\over {\cal V}^2} \\[0.1cm]
     {O(1)\, \over {\cal V}^2} &
      O(1)
 \end{matrix}\right)\, .
}
Computing the eigenvalues of the matrix $({\cal K}^{-1})^{am} V_{mb}$ we
obtain
\eq{
\label{massev}
      M_1\simeq {M_{\rm pl}\over {\cal V}}\, ,\hspace{40pt}
      M_2\simeq {M_{\rm pl}\, \sqrt{\lambda_f}\over {\cal V}^{3/2}}\,.
}
Diagonalizing then the axion sector, we find the relation between the
old axions and the new eigensystem as 
\eq{
\label{eigenveca}
    \phi^{(1)}&\simeq O(1)\, {\cal V}^{1\over 2} \psi^{(1)} + O(1)\,
    {\lambda_f} \psi^{(2)} \,,\\
    \phi^{(2)}&\simeq  O(1)\,\frac{\lambda_f}{{\cal V}^{1\over 2}}
     \psi^{(1)} + O(1)\,  \psi^{(2)}
\, .
}
Therefore, $\phi^{(1)}=a_s$ is mostly $\psi^{(1)}$ and $\phi^{(2)}=\theta$ is mostly
$\psi^{(2)}$. These relations will become important in the discussion
of reheating in section \ref{sec_reheat}.

Before closing this section, let us comment on the validity of the regime $f>M_{\rm pl}$.
We have seen that $f>M_{\rm pl}$ leads to $g_s>1$, so that
one is actually outside the perturbative regime where
the effective field theory was computed for. This clearly is the
most severe issue of the natural inflation model presented in
this letter. Let us make two comments:
\begin{itemize}
\item{The perturbative axionic couplings are protected by a shift symmetry,
and hence the main concern is about the description prior to the stabilization
of the saxionic directions.}
\item{The actual order parameter in the LVS is ${\cal V}^{-1}$,
so it is not completely obvious that for say $1<g_s<10$ we 
immediately get  unstable  results. }
\end{itemize}
Concerning the last point, let us have a look at the 
corrections to the scalar potential originating from
 one-loop corrections to the K\"ahler potential.
In \cite{Berg:2007wt} an extended  no-scale structure was observed, for which the 
corrections to the scalar potential become
\eq{
         V&=V_{LVS}+V_{{\rm 1-loop}}\\
   &\simeq \hphantom{+}{\sqrt{\tau_s} e^{-2\alpha_s \tau_s}\over {\cal V}}
   -{{\tau_s} e^{-\alpha_s \tau_s} \over {\cal V}^2}+
     {1\over g_s^{3/2} {\cal V}^3} \\
   &\hspace{9pt} +{g_s^2\over
           {\cal V}^{3} \sqrt{\tau_b}} + {g_s^2\over
           {\cal V}^{3} \sqrt{\tau_s}}\,.
}
In the LVS minimum these terms scale as
\eq{
         V\simeq {1\over g_s^{3\over 2} {\cal V}^3} + {g_s^2\over
           {\cal V}^{10\over 3}} + {g_s^2\over
           {\cal V}^{3} \sqrt{\log \left(g_s^{1/2}\,{\cal V}\right)}}\, .
}
Therefore, higher orders in $g_s$ are accompanied by further volume
suppression factors. Of course, we cannot control all higher loop-corrections,
but as long as they still come with suppressions in ${\cal V}$,  the LVS
scalar potential could still be trusted for $g_s>1$ and ${\cal V}$
sufficiently large.


\section{Universal axion monodromy inflation}

One approach to avoid the regime $f>M_{\rm pl}$ is to consider models
where the axion shift symmetry is broken either by D-branes or by fluxes.
Here, we  want to construct  such a model for the universal
axion, whose   shift symmetry is broken by the three-form flux.
In this case a non-oscillatory potential is generated, allowing
the axion to roll over trans-Planckian distances in field space without
needing $\theta/f<1$. Thus, the periodic axion unwraps to allow
for non-trivial monodromies.

We consider a model in the flux landscape, where the fluxes 
break the shift symmetry of the axion slightly by giving it 
a parametrically small mass 
\eq{
      m_\theta\simeq {\lambda_0\, M_{\rm pl} \over {\cal V}} \,.
}
Again, we assume that the complex-structure moduli
and the dilaton can be fixed at a hierarchically bigger mass scale
by a combination of fluxes and  contributions
to the scalar potential violating its no-scale 
structure \cite{Blumenhagen:future}.
In this case, the value of the superpotential in the minimum will
have the following simple form
\eq{
         W_0=w_0 + \lambda_0 \,\theta\, ,
} 
guaranteeing that for the axion becoming massless the shift symmetry
of the superpotential is restored.
For the mass of axion to be smaller than the masses of K\"ahler 
moduli we require $\lambda_0\ll {\cal V}_0^{-{1\over 2}}$.
The effective potential for the axion can then be written as
\eq{
  V_{\rm eff}\simeq  {\lambda_0^2 \over {\cal V}^2}\, \theta ^2 \, .
}
Interestingly,  due to the flux-breaking of the axionic shift symmetry
we eventually get an effective axion potential which is 
of quadratic order and therefore can be considered a candidate for  
a stringy realization of chaotic inflation. Moreover, this model
is of the type of axion monodromy inflation, as the compact interval 
$\theta=\theta+ 2\pi f$ gets unwrapped.

For computing the masses and the mixing of 
the two axions $\phi^{(1)}=a_s$ and $\phi^{(2)}=\theta$,
we now have the mass matrix
\eq{
         V_{ab}=\left(\begin{matrix} {O(1)\over {\cal V}^3} &
     0 \\[0.1cm]
     0 &
      {O(1)\, \lambda_0^2 \over {\cal V}^2}
 \end{matrix}\right) \, .
}
For the  eigenvalues of the matrix ${\cal K}^{-1}_{am} V^m{}_{b}$ we
find 
\eq{
  \label{massev_am}
       M_1\simeq {M_{\rm pl}\over {\cal V}}\, ,\hspace{40pt}
      M_2\simeq {M_{\rm pl}\, \lambda_0\over {\cal V}}\,,
}
with eigenvectors
\eq{
\label{eigenvecb}
    \phi^{(1)}&\simeq O(1)\, {\cal V}^{1\over 2} \psi^{(1)} + O(1)\,
    {\lambda_0^2\over {\cal V}} \psi^{(2)} \,,\\
    \phi^{(2)}&\simeq {O(1)\over {\cal V}^{3\over 2}} \psi^{(1)} + O(1)\,  \psi^{(2)}
\, .
}
Therefore, as for universal natural inflation,
$\phi^{(1)}=a_s$ is mostly $\psi^{(1)}$ and $\phi^{(2)}=\theta$ is mostly
$\psi^{(2)}$.

It is important to know how stable this model is in the UV complete
theory, i.e. whether it suffers for instance from an $\eta$-problem
due to other corrections depending on the inflaton.
Shift symmetry breaking  effects should either be proportional to 
the fluxes or come from non-perturbative effects 
like $E3$-brane instantons. 
The effect of the latter should be harmless as they  are 
exponentially suppressed
relative to the leading instanton contribution to $W$.
In string theory,  higher-order flux induced corrections 
to the K\"ahler potential  are not well understood
so that  their analysis is beyond the scope of this letter.
One should keep in mind that they could be potentially
dangerous.

\section{Reheating}
\label{sec_reheat}

At the end of inflation the axion $\theta$ oscillates around its minimum and
decays into the various modes it couples to, thereby reheating the universe. 
It is often a difficult problem to guarantee that the inflaton
mostly decays into the visible Standard Model sector \cite{Cicoli:2010ha, Cicoli:2010yj}. Too many 
decays into the hidden sectors produces dark matter and
can over-close the universe. 
In order to estimate
this, we consider the coupling of the inflaton to gauge fields localized
on intersecting $D7$-branes.

Recall that in concrete  string model building one has to satisfy the $D7$- and
$D3$-brane tadpole cancellation 
conditions \cite{Collinucci:2008pf,Blumenhagen:2008zz,Plauschinn:2008yd}.  Say we have stacks 
of in general magnetized $D7$-branes wrapping divisors $D_a$ in the
internal Calabi-Yau manifold. Then the $D7$-brane tadpole cancellation
condition is just a homology condition
\eq{
\label{seventad}
    \sum_a N_a ( [D_a]+[D'_a] )=8\, D_{\rm O7} \,,
}
where $D_a'$ denotes the orientifold image of $D_a$.
The $D3$-brane tadpole condition reads as follows
\eq{
\label{threetad}
  N_{\rm D3} + {N_{\rm flux}\over 2} + N_{\rm gauge}=
    {N_{\rm O3}\over
    4}+
    {\chi(D_{\rm O7})\over 12}+
    \sum_a N_a {\chi(D_{a})\over 24} ,
}
where $N_{\rm D3}$ are the number of $D3$-branes, $N_{\rm flux}$ is a contribution
from the $H\wedge F$-flux, $N_{\rm gauge}$ is the contribution
from the magnetic flux on the $D7$-branes and $\chi(D_a)$ is the Euler
characteristic of the complex surface $D_a$.

For simplicity, we first  assume that
the $D7$-branes wrap the small cycle, even though this is against the instanton zero-mode arguments of \cite{Blumenhagen:2007sm}. We come back to this point below. In order to obtain chirality,
the Standard-Model (SM) $D7$-branes are also equipped with a non-vanishing
background gauge flux $f$. Therefore, the axionic 
coupling to a SM brane becomes
\eq{
       {\cal L}_{\rm SM} \simeq {1\over M_{\rm pl}} \Big(a_s + h(f)\, \theta\Big)\, 
F\wedge F\,\Bigr|_{\rm SM} \,.
}  
To satisfy the $D7$-brane tadpole cancellation con\-di\-tions,
generically  additional hidden $D7$-branes have to be present. However, these
branes can usually be chosen non-chiral so that they just serve as ``filler'' branes.
Their axionic couplings are therefore
\eq{
       {\cal L}_{\rm hid} \simeq {1\over M_{\rm pl}} \, a_s\, F\wedge
         F \,\Bigr|_{\rm hid} \,.
}  
Thus, there is a clear distinction in couplings to the axions between these two kinds of $D7$-branes.
Note that if there are also $D3$-branes present, they would couple to the
axion  in a direct way
\eq{
       {\cal L}_{\rm D3} \simeq {1\over M_{\rm pl}} \theta\, 
F\wedge F \,\Bigr|_{D3}\, .
}


\subsection*{Reheating for natural  inflation}

For natural inflation, we recall from equation \eqref{eigenveca} that the inflaton is mostly $\psi^{(2)}$
which couples to the SM sector dominantly as
\eq{
      {\cal L}_{\rm SM} \simeq {1\over M_{\rm pl}} h(f_{SM})\, \psi^{(2)}\, 
F\wedge F\,\Bigr|_{\rm SM} \,,
}
while it couples to the hidden sector as
\eq{
       {\cal L}_{\rm hid} \simeq {1\over M_{\rm pl}}{\lambda_f} \, \psi^{(2)}\, F\wedge
         F\,\Bigr|_{\rm hid} \,.
}  
Assuming that we can saturate the $D3$-brane tadpole condition \eqref{threetad} by the 
flux contributions $N_{\rm flux}$ and $N_{\rm gauge}$, i.e. without $D3$ branes, the hidden sector decays are suppressed relative to SM decays as
\eq{
\label{reheatsup}
         {   \Gamma(\theta\to \gamma^2_{\rm hid})\over
             \Gamma(\theta\to \gamma^2_{\rm SM})}=\lambda_f^2
             \,,
} 
and reheating predominantly occurs into the SM degrees of freedom.
For the decay rate induced by a dimension-five operator ${g\over M_{\rm pl}}\,\theta\, F\wedge F$,
here we have used the formula
\eq{
   \Gamma(\theta\to \gamma^2)={g^2\, m_\theta^3\over 64\pi\,M_{\rm pl}^2} \, .
}

However, in \cite{Blumenhagen:2007sm} the authors pointed out that there exist a tension between
the chirality of intersecting $D7$-branes and stabilizing
the corresponding four-cycles via $E3$-brane instantons.
This can be reconciled by sequestering the SM and moduli-stabilization sectors,
that is placing the chiral Standard Model on a different 
del-Pezzo surface. Such a scenario was analyzed in detail in 
\cite{Blumenhagen:2009gk}. 
In this case, the complex K\"ahler moduli controlling
 the SM sector are stabilized via D-terms and the generalized
Green-Schwarz mechanism. Performing the same computation
as before with
\eq{
         V_{ab}=\left(\begin{matrix} {O(1)\over {\cal V}^2} &
      0 \\[0.1cm]
     0 &
      {O(1)\, \lambda_f \over {\cal V}^3}
 \end{matrix}\right) \,,
}
one finds 
\eq{
    \phi^{(1)}&\simeq O(1)\, {\cal V}^{\frac{1}{2}}\,  \psi^{(1)} + O(1)\,
    {\lambda_f\over {\cal V}^3} \psi^{(2)} \,,\\
    \phi^{(2)}&\simeq {O(1)\over {\cal V}^{\frac32}} \psi^{(1)} + O(1)\,  \psi^{(2)}
\, .
}
Therefore, compared to \eqref{reheatsup} the hidden sector decays are further volume 
suppressed 
\eq{
\label{reheatsupb}
         {   \Gamma(\theta\to \gamma^2_{\rm hid})\over
             \Gamma(\theta\to \gamma^2_{\rm SM})}=\left({\lambda_f\over {\cal V}^3}\right)^2 \,.
}


\subsection*{Reheating for axion monodromy inflation}

Performing the same analysis  for our model of axion-monodromy
inflation using  \eqref{eigenvecb}, we find that
the hidden sector decays are also suppressed relative to SM decays 
\eq{
\label{reheatsupmono}
         {   \Gamma(\theta\to \gamma^2_{\rm hid})\over
             \Gamma(\theta\to \gamma^2_{\rm SM})}={\lambda^4_0\over {\cal
             V}^2}\, .
} 
Localizing the SM on  a sequestered $D7$-brane, the suppression
is even stronger 
 \eq{
\label{reheatsupbmono}
         {   \Gamma(\theta\to \gamma^2_{\rm hid})\over
             \Gamma(\theta\to \gamma^2_{\rm SM})}={\lambda^4_0\over {\cal V}^4} \,.
} 
Thus, both  universal natural inflation and universal axion monodromy
inflation provide a natural mechanism for reheating mainly into SM degrees of
freedom. The reheating temperature comes out as
\eq{
            T_R\simeq \sqrt{\Gamma M_{\rm pl}}\simeq h(f_{SM})\, \sqrt{
              m_\theta^3\over M_{\rm pl}}\simeq 10^{10}{\rm GeV} \, ,
}
which is much higher than the big bang nucleosynthesis temperature
$T_{\rm BBN}\sim 1\,$MeV.


\subsection*{Inflationary scales}
 
Finally, let us estimate the relevant scales.
Choosing as a reference value $\lambda_0=4\cdot 10^{-3}$ 
and comparing the inflaton
mass $m_\theta\simeq 10^{13}$GeV  with 
$M_2$ of our sequestered axion monodromy inflation model in \eqref{massev_am},
we find for the volume ${\cal V}=580$. The suppression factor
in \eqref{reheatsupbmono} is then $(\lambda_0/{\cal V})^4\simeq 10^{-21}$.
The string scale $M_s\simeq {M_{\rm
  pl}\over \sqrt{\cal V}}$ comes out as $M_s\simeq 10^{17}$GeV and
the gravitino mass $M_{3/2}={M_{\rm
  pl}\over {\cal V}}$ as $M_{3/2}\simeq 4\cdot 10^{15}$GeV.
In the sequestered LVS scenario \cite{Blumenhagen:2009gk}, the soft terms are suppressed
as
\eq{
           m_{\rm soft}\simeq {M_{\rm pl}\over {\cal V}^2}\simeq 7\cdot 10^{12}{\rm
             GeV} \,,
}
pointing to an intermediate  scale of supersymmetry breaking, as it was also 
found recently in \cite{Ibanez:2014zsa}.


\section{Conclusions}

In this letter we have proposed a type IIB, LVS-like string-realization of
both natural  inflation and axion monodromy inflation, 
where the role of the inflaton is played
by the universal axion, whose scalar partner is the dilaton.

Concerning natural inflation, under the assumption that the universal 
axion is not already fixed
by a leading-order flux-induced potential, i.e. its shift symmetry is
still intact, we have argued
that a non-perturbative contribution to the superpotential
coming from a magnetized $E3$-brane instanton gives rise to a  leading-order
potential. The resulting mass of the axion turned out to be smaller than
that of the K\"ahler moduli and than the small-cycle axion. 

Concerning axion monodromy inflation,  we have argued
that shift-symmetry breaking fluxes can still allow for 
an  axion of parametrically small mass.
Interestingly, this shift-symmetry breaking potential
could  be quadratic in the axion, thus providing
a string-derived candidate  of chaotic inflation. 

In both cases, the big cycle
axion was still massless at this stage and could still lead
to dark radiation \cite{Cicoli:2012aq,Higaki:2012ar} via the large K\"ahler modulus decay.

As one of main results, we have  described a natural mechanism  
guaranteeing that at the end
of inflation the inflaton
predominantly decays into SM particles. This is achieved
by having only the SM branes carry chirality-inducing gauge flux,
which leads to a direct coupling of the inflaton to the 
SM degrees of freedom. Note that this   mechanism works for both natural
and axion monodromy inflation. 

The predictions of universal axion  inflation
could be met by choosing the overall volume $\mathcal V$ 
of the compactification space to be of the order $10^2-10^3$. This
led to soft masses in the sequestered LVS of the order
of $10^{12}$GeV, implying a high-scale susy breaking.
The other moduli are very heavy in this model, so that there
is no cosmological moduli problem.

Having the axion decay constant larger than the Planck scale 
required the string coupling constant to be larger than one.
We have presented one argument why this (F-theoretic) regime
might still be under control in the LVS. This is of course
the weakest point of our model;  nevertheless,
we think that it shows some new and interesting features. 
Moreover, 
in this letter we have simply assumed that the dilaton can be stabilized
by either fluxes or no-scale breaking effects such that its mass is 
hierarchically bigger than the  axion mass scale. 
However, the corresponding stabilization mechanism of the axio-dilaton
deserves a closer technical investigation \cite{Blumenhagen:future}.


\subsubsection*{Acknowledgments}
\vskip0.5em

\noindent
We thank T. Grimm for useful discussion.
E.P. is supported by the MIUR grant  FIRB RBFR10QS5J
and by the COST Action MP1210. He furthermore wants to
thank the Max-Planck-Institute for Physics in Munich for kind hospitality.


\subsection*{References}
\vskip0.5em

\bibliographystyle{elsarticle-num}
\bibliography{references}


\end{document}